\documentclass[12pt]{article}
          \usepackage{amsfonts}
          \usepackage{amssymb}

\usepackage[margin=1in]{geometry}
\newtheorem{thm}{Theorem}

\newtheorem{defn}{Definition}

\def\tr{\hbox{Tr}}

\def\be{\begin{eqnarray}}
\def\ee{\end{eqnarray}}
\def\bee{\begin{eqnarray*}}
\def\eee{\end{eqnarray*}}
\def\bmx{\begin{pmatrix}}
\def\emx{\end{pmatrix}}

\def\ts{\textstyle}
\def\bra{\langle}
\def\ket{\rangle}

\def\rt2{\ts \frac{1}{\sqrt{2}} }

\def\ot{\otimes}

\title{Potential output purity of completely positive maps}

    \author{Christopher King\\
    \\
{\small      Department of Mathematics} \\
{\small      Northeastern University} \\
{\small      Boston MA 02115}}

\begin{document}

\maketitle
     
\abstract{The notion of potential output purity of a completely positive map is introduced as a generalization of the regularized output purity.
An upper bound is derived for this quantity, and for 
several classes of maps (including CQ, QC and Hadamard channels) it is shown that  potential purity does not exceed the standard output purity.
As an application the potential purity is used to bound the logarithmic Sobolev constant of a product of depolarizing channel semigroups.}

\section{Introduction}
Quantum entanglement is a valuable resource for processing information, and in many cases 
entanglement is crucial for producing a gap between the optimal performance
which can be achieved using classical resources and the optimum using quantum resources. 
One prominent example where this gap arises is the transmission of classical and quantum information through a channel, 
where entangled inputs, entangled measurements, prior entanglement between sender and receiver and so on 
have been shown to enhance the capacity.
In this paper we will use the output purity $\| {\cal N} \|_{q \rightarrow p}$ of a channel ${\cal N}$
to investigate the effects of entangled input states.
Here $q,p \ge 1$ are used to define Schatten norms on the input and output
spaces respectively, and $\| {\cal N} \|_{q \rightarrow p}$ is the norm of ${\cal N}$ viewed as 
a map between these Banach spaces. 
When $q=1$ the output purity is closely related to the minimal output Renyi entropy of the channel.
It is known that the minimal output Renyi entropy may be non-additive for product channels
when entangled input states are used. In this paper we will make use of output purity to
explore this non-additivity for product channels.

\medskip
We will work with a quantity which measures the contribution of entanglement to the output purity of a product channel,
or more generally a product of completely positive maps. This quantity, which we call
{\em potential output purity}, is inspired by the notion of potential capacity described below. The basic idea
is to capture the amount by which the output purity of a product map can differ from the product of output
purities of the component maps. Alternatively the potential output purity measures the effectiveness of the map
to catalyze output purity when it is used in a product.

\medskip
The notion of potential capacity appeared in the paper of
Smith, Smolin and Winter \cite{SSW2008}, who introduced the value added quantum capacity of a channel
as a way to measure the maximal violation of additivity which can be `catalyzed' by the channel. This notion was later extended by
Winter and Yang \cite{WY2016} to apply to any capacity-like quantity $C$: the potential capacity $C^{(pot)}$ of a channel ${\cal N}$ is
\be\label{def:WY-pot}
C^{(pot)}({\cal N}) = \sup_{\cal M} \left[ C({\cal N} \ot {\cal M}) - C({\cal M}) \right]
\ee
where the supremum on the right side runs over all channels ${\cal M}$.
If the quantity $C({\cal N} \ot {\cal M})$ is additive when restricted to product states,
and non-additive for some entangled states, then $C({\cal N} \ot {\cal M}) - C({\cal M})$
measures the amount of `non-classical' capacity which is available
in the product channel ${\cal N} \ot {\cal M}$. In this sense the potential capacity provides a way
to measure the gap between the optimal capacity which can be attained using classical states and entangled states.

\medskip
Motivated by this notion, in Section \ref{sect2} 
we define the potential output purity $\| \Phi \|_{q \rightarrow p}^{(pot)}$ of a completely positive map.
We present some general properties of potential purity, and describe its relation to the regularized output purity.
Then in Section \ref{sect3} we state our main results:
Theorems \ref{thm1} and \ref{thm2}  provide upper bounds for the potential purity, and Theorems \ref{thm3} and \ref{thm4} present some special cases
where $\| \Phi \|_{q \rightarrow p}^{(pot)} = \| \Phi \|_{q \rightarrow p}$. 
Sections \ref{sect4}, \ref{sect5} and \ref{sect6} contain the proofs of these results.
Finally in Section \ref{sect7} we discuss some related results and open questions.

\section{Notation and definitions}\label{sect2}
\subsection{Notation}
We denote by ${\cal M}_d$ the algebra of $d \times d$ (complex-valued) matrices,
and ${\cal M}_d^{+} \subset {\cal M}_d$ the cone of positive semidefinite matrices.
For all $p \ge 1$ the Schatten norm of $A \in {\cal M}_d$ is defined as
\be\label{def:Schatten}
\| A \|_p = \left( \tr |A|^p \right)^{1/p}, \quad |A| = (A A^* )^{1/2}
\ee
and for a map $L : {\cal M}_d \rightarrow {\cal M}_{d'}$ we define the corresponding operator norms
\be\label{def:map-norm}
\| L \|_{q \rightarrow p} = \sup_{A \in {\cal M}_d} \, \frac{\| L(A) \|_p}{\| A \|_q}, \quad q,p \ge 1
\ee
The adjoint of $L$ is the map $L^* : {\cal M}_{d'} \rightarrow {\cal M}_{d}$ defined by the relations
\be\label{def: L^*}
\tr \, A \, L(B)  = \tr \, L^*(A) \, B \quad \mbox{for all $A \in {\cal M}_{d'}$, $B \in {\cal M}_{d}$}
\ee

\medskip
We write $I_d$ for the identity matrix in ${\cal M}_d$, and ${\it id}_d$ for the identity map on
${\cal M}_d \rightarrow {\cal M}_{d}$. 
The superoperator $L : {\cal M}_d \rightarrow {\cal M}_{d'}$ is completely positive (CP) if 
${\it id}_n \ot L$ is positivity preserving on ${\cal M}_n \ot {\cal M}_d$ for all $n \ge 1$.
A quantum channel is a completely positive trace preserving map. For a completely positive map $\Phi$
the quantity $\| \Phi \|_{q \rightarrow p}$ will be referred to as output purity of the map, and
for all $q,p \ge 1$ we have \cite{Wat2005}, \cite{Au1}
\be\label{purity-CP}
\mbox{$\Phi$ is CP $\Rightarrow$} \,\, \| \Phi \|_{q \rightarrow p} = \sup_{A \in {\cal M}_d} \, \frac{\| \Phi(A) \|_p}{\| A \|_q}
= \sup_{A \in {\cal M}_d^{+}} \, \frac{\| \Phi(A) \|_p}{\| A \|_q}
\ee

\medskip
For a map $L : {\cal M}_d \rightarrow {\cal M}_{d'}$ the Choi-Jamiolkowski  matrix is 
the element $X_{L} \in {\cal M}_d \ot {\cal M}_{d'}$ defined by
\be\label{def:Choi}
X_{L} = \sum_{i,j=1}^d | i \ket \bra j | \ot L(| i \ket \bra j |)
\ee
where $\{ | i \ket \}$ is an orthonormal basis of $\mathbb{C}^d$.

\medskip
A completely positive map ${\cal E} : {\cal M}_d \rightarrow {\cal M}_{d'}$ is called {\em entanglement breaking} if it can be written in the form
\be\label{def:E-B}
{\cal E}(A) = \sum_{k=1}^N \tr (X_k A) \, R_k
\ee
where $\{X_k \in {\cal M}_d^{+}\}$ and $\{R_k \in {\cal M}_{d'}^{+}\}$ are collections of positive semidefinite matrices.
The map ${\cal E}$ is CQ if $N=d$ and $X_k = | k \ket \bra k |$, $k=1,\dots,d$; ${\cal E}$ is QC if $N = d'$ and
$R_k = | k \ket \bra k |$, $k=1,\dots,d'$. Note that ${\cal E}$ is CQ if and only if ${\cal E}^*$ is QC.

\medskip
A completely positive map ${\cal E} : {\cal M}_d \rightarrow {\cal M}_{d}$ is called a {\em Hadamard map}
if it acts by Hadamard product with a matrix $C \in {\cal M}_d^{+}$. That is, 
for some choice of basis $\{ | i \ket \} \in \mathbb{C}^d$ and
for any $A = \sum_{i,j=1}^d a_{ij} | i \ket \bra j |$
we have
\be\label{def:Hada}
{\cal E}(A) =  \sum_{i,j=1}^d c_{ij} \, a_{ij} | i \ket \bra j |
\ee
where $C = \sum_{i,j=1}^d c_{ij} | i \ket \bra j |$. In this case we will also write ${\cal E} = H_C$.
Note that in the special case where $c_{ij}=1$ for all $i,j=1,\dots,d$ we have
$H_C={\it id}_d$, the identity map.

\subsection{Potential output purity}
Smith, Smolin and Winter \cite{SSW2008} introduced the notion of value added quantum capacity of a channel $\cal N$,
given by the expression $\sup_{\cal E} \left( Q({\cal E} \ot {\cal N}) - Q({\cal E}) \right)$ where $Q$ is the quantum capacity, and
the supremum runs over all quantum channels.
This notion was revisited by Winter and Yang \cite{WY2016} who coined the term potential capacity, and explored the
notion for a variety of other capacities. We will borrow this terminology to define a corresponding notion for
output purity.

\begin{defn}\label{def1}
Let $\Phi$ be a completely positive map on ${\cal M}_d$. 
The potential output purity is
\be\label{def:pot-pur}
\| \Phi \|_{q \rightarrow p}^{(pot)} = \sup_n \, \sup_{{\Omega_n}} \,
\frac{\| \Phi \ot \Omega_n \|_{q \rightarrow p}}{\| \Omega_n \|_{q \rightarrow p}}, \quad q,p \ge 1
\ee
where the inside supremum runs over all completely positive maps $\Omega_n$ on ${\cal M}_n$,
and the outside supremum runs over all dimensions $n \ge 1$.
\end{defn}

\medskip
\par\noindent{\it Remarks:} our investigation of the quantity (\ref{def:pot-pur}) is motiovated by the 
known violations of multiplicativity of maximal output purity for the case $q=1$, $p > 1$ \cite{HaydenWinter, ASW}.  That is,
it is known that for all $p>1$ there are high-dimensional quantum channels for which
\be
\| \Phi \ot \overline{\Phi}\|_{1 \rightarrow p} > \| \Phi \|_{1 \rightarrow p}^2
\ee
where $\overline{\Phi}$ is the complex conjugate channel. For such channels there is a gap between
the output purity and the potential output purity, that is
\be\label{pot>pur}
\| \Phi \|_{1 \rightarrow p}^{(pot)} > \| \Phi \|_{1 \rightarrow p}
\ee
In fact as the dimension increases it becomes almost certain that a randomly selected channel will satisfy (\ref{pot>pur})
(however explicit examples of such channels are known only for $p > 2$ \cite{GrHoPan}). Using continuity it follows that there must also be a gap
between purity and potential purity for $q = 1 + \epsilon$ for $ \epsilon$ sufficiently small. It seems plausible that this gap persists for all $q > 1$,
and this motivates the results in the next section where we establish uniform bounds on the
size of the potential output purity. We note in passing that in general for $q > 1$ the supremum in the norm
$\| \Phi \|_{q \rightarrow p} = \sup_A \| \Phi(A) \|_p \, \| A \|_q^{-1}$ is not achieved on a rank one matrix $A$, and this eliminates some of 
the nice methods which have been developed for analyzing the case $q=1$.

\medskip
We note a few properties of the potential purity which follow immediately from its definition. First, it is clear that
\be\label{pot-ineq0}
\| \Phi \|_{q \rightarrow p}^{(pot)} \ge \| \Phi \|_{q \rightarrow p}
\ee
Second, the potential purity is sub-multiplicative: for any maps $\Phi_1$, $\Phi_2$ and $\Omega$ we have
\be
\frac{\| \Phi_1 \ot \Phi_2 \ot \Omega \|_{q \rightarrow p}}{\| \Omega \|_{q \rightarrow p}} =
\frac{\| \Phi_1 \ot \Phi_2 \ot \Omega \|_{q \rightarrow p}}{\| \Phi_2 \ot \Omega \|_{q \rightarrow p}} \,
\frac{\| \Phi_2 \ot \Omega \|_{q \rightarrow p}}{\| \Omega \|_{q \rightarrow p}} 
 \le 
\| \Phi_1 \|_{q \rightarrow p}^{(pot)} \, \| \Phi_2 \|_{q \rightarrow p}^{(pot)}
\ee
and since this holds for all $\Omega$ it follows that
\be
\| \Phi_1 \ot \Phi_2 \|_{q \rightarrow p}^{(pot)}  \le
\| \Phi_1 \|_{q \rightarrow p}^{(pot)} \, \| \Phi_2 \|_{q \rightarrow p}^{(pot)}
\ee
We thus get the following sequence of inequalities for any completely positive map $\Phi$:
\be\label{ineq-seq}
\| \Phi \|_{q \rightarrow p} \le \lim_{n \rightarrow \infty} \left(\| \Phi^{\ot n} \|_{q \rightarrow p} \right)^{1/n}
\equiv \| \Phi \|_{q \rightarrow p}^{(reg)}
\le \lim_{n \rightarrow \infty} \left(\| \Phi^{\ot n} \|_{q \rightarrow p}^{(pot)} \right)^{1/n}
\le \| \Phi \|_{q \rightarrow p}^{(pot)}
\ee
where $\| \Phi \|_{q \rightarrow p}^{(reg)}$ is the regularized purity.

\medskip
Third, it is known that the output purity is multiplicative for all $q \ge p$ \cite{DJKR}.
That is, for all completely positive maps $\Phi$ and $\Omega$, and all $q \ge p \ge 1$,
it is known that
\be
\| \Phi \ot \Omega \|_{q \rightarrow p} = \| \Phi \|_{q \rightarrow p} \, \| \Omega \|_{q \rightarrow p}
\ee
and thus 
\be\label{pot-eq}
\| \Phi \|_{q \rightarrow p}^{(pot)} = \| \Phi \|_{q \rightarrow p} \quad \mbox{for all $q \ge p \ge 1$}
\ee
In view of this we will restrict attention to the case $q < p$ in the following results.

\section{Results}\label{sect3}
Our first result is a bound on the potential purity which applies for all maps and for all
$q,p \ge 1$. 
Recall the notation $X_{\Phi}$ for the Choi matrix of the map $\Phi$ as defined in (\ref{def:Choi}).

\begin{thm}\label{thm1}
Let $\Phi : {\cal M}_d \rightarrow {\cal M}_{d'}$ be a completely positive map and $q,p \ge 1$. Then
\be\label{thm1:eq1}
\| \Phi \|_{q \rightarrow p}^{(pot)} \le \alpha(q,p) \, \| X_{\Phi} \|_2
\ee
where
\be\label{thm1:eq2}
\alpha(q,p) = \begin{cases}{ 1 & for $q \le 2 \le p$ \cr d^{1-2/q} & for $2 < q < p$ \cr (d')^{2/p -1} & for $q < p < 2$ }\end{cases}
\ee
\end{thm}

\medskip
The next result provides a potentially tighter bound for the case where $q \le 2 \le p$.

\begin{thm}\label{thm2}
Let $\Phi : {\cal M}_d \rightarrow {\cal M}_{d'}$ be a completely positive map. Then for
all $1 \le q \le 2 \le p$,
\be\label{thm2:eq1}
\| \Phi \|_{q \rightarrow p}^{(pot)} \le  \min \{\| \Phi \|_{p \rightarrow p}, \, \| X_{\Phi} \|_2 \}
\ee
\end{thm}

\medskip
\par\noindent{\it Remarks:} the right side of (\ref{thm2:eq1}) may be achieved with either term.
For example, if $\Phi$ is a unital trace-preserving channel then $\| \Phi \|_{p \rightarrow p} \le 1$
for all $p \ge 1$ \cite{PWPR}, while we also have 
\be
\| X_{\Phi} \|_2^2 \ge \sum_{j=1}^d \| \Phi(| j \ket \bra j |) \|_2^2
\ge d \, \| d^{-1} I_d \|_2^2 \ge 1
\ee
and so $\| \Phi \|_{p \rightarrow p} \le \| X_{\Phi} \|_2$ in this case.
Conversely, for the channel $T : {\cal M}_d \rightarrow {\cal M}_1$, $T(A) = \tr (A)$, we have
$\| T \|_{p \rightarrow p} = d^{1 - 1/p}$ while
\be
X_{T} = \sum_{i,j=1}^d | i \ket \bra j | \ot T(| i \ket \bra j |) = I_d
\ee
and thus $\| X_{T} \|_2 = d^{1/2}$, so in this case $\| T \|_{p \rightarrow p} > \| X_{T} \|_2$ for all $p > 2$.

\medskip
Theorems \ref{thm1} and \ref{thm2} will be proved in the next section. 
Our next two results present some special cases where there is no gap between potential purity and purity,
that is where equality holds in (\ref{pot-ineq0}).
In the paper \cite{King-EB} it was shown that equality in (\ref{pot-ineq0}) holds at $q=2$ 
for a class of entanglement-breaking maps which includes the CQ maps.
In Theorem \ref{thm3} we extend this result to all $q,p \ge 1$ for CQ and QC maps.

\begin{thm}\label{thm3}
The equality $\| \Phi \|_{q \rightarrow p}^{(pot)} = \| \Phi \|_{q \rightarrow p}$ holds 
for all $q,p \ge 1$ when $\Phi$ is a CQ or QC map. 
\end{thm}

\medskip
For the range $1 \le q \le 2 \le p$, Watrous showed that equality in (\ref{pot-ineq0}) holds when $\Phi$ is the identity map \cite{Wat2005}.
Our next Theorem extends Watrous' result to include all Hadamard maps
(recall that the identity map is one example of a Hadamard map).

\begin{thm}\label{thm4}
The equality $\| H_C \|_{q \rightarrow p}^{(pot)} = \| H_C \|_{q \rightarrow p}$ holds 
for all Hadamard maps $H_C$ for the range of parameters $1 \le q \le 2 \le p$.
\end{thm}

\medskip
\par\noindent{\it Remarks:} 
Equality in (\ref{pot-ineq0}) has been shown for unital qubit channels \cite{King-HC} in the range $1 \le q \le 2 \le p$. 
Equality  is also known to hold
for maps with entrywise non-negative Choi matrices when $q=2$ and $p$ is an even integer \cite{KNR}.
Except for Theorem \ref{thm3} not much is known about
conditions for equality outside the range of values $1 \le q \le 2 \le p$.
Hypercontractivity bounds can provide some related but weaker results  \cite{King-HC}: for example, for
the qubit depolarizing channel $\Delta_{\lambda}$ we have
$\| \Delta_{\lambda}^{\ot n} \|_{q \rightarrow p} = \| \Delta_{\lambda} \|_{q \rightarrow p}^n$
when $0 \le \lambda \le 1$, $q \ge 2$, $p \le 1 + (q-1) \lambda^{-2}$.

\section{Proof of Theorems \ref{thm1} and \ref{thm2}}\label{sect4}
\subsection{Proof of Theorem \ref{thm1}}
Let $\Omega: {\cal M}_n \rightarrow {\cal M}_{n'}$ be a completely positive map. Using H\"older duality we have 
\be\label{def:Holder-dual}
\| \Phi \ot \Omega \|_{q \rightarrow p} = \sup_{B \in {\cal M}_{d' n'}} \, \sup_{A \in {\cal M}_{d n}} \,
\frac{\tr \Big[ B^* (\Phi \ot \Omega)(A) \Big]}{\| B \|_{p'} \, \| A \|_q}
\ee
where $p'$ is the conjugate value to $p$, defined by
\be\label{def:p'}
\frac{1}{p'} = 1 - \frac{1}{p}
\ee
We will prove Theorem \ref{thm1} by deriving a bound for $\tr \Big[ B^* (\Phi \ot \Omega)(A) \Big]$ and applying (\ref{def:Holder-dual}).

\medskip
We first introduce a new notation for the entries of the Choi matrix of $\Phi$ (recall the definition of $X_{\Phi}$ from
(\ref{def:Choi})):
\be\label{def:C}
C_{ij, kl} =  \tr \Big( \Big[ | j \ket \bra i | \ot | l \ket \bra k | \Big] X_{\Phi} \Big), \quad
i,j=1,\dots,d; \,\, k,l =1,\dots, d'
\ee
With this notation we have
\be
\Phi(| i \ket \bra j |) = \sum_{k,l=1}^{d'} C_{ij, kl} \, | k \ket \bra l |, \quad
X_{\Phi} = \sum_{i,j=1}^d  \sum_{k,l=1}^{d'} C_{ij, kl} \, | i \ket \bra j | \ot | k \ket \bra l |
\ee
In particular this provides an expression for $\| X_{\Phi} \|_2$ which we will use later:
\be\label{cal-L2}
\| X_{\Phi} \|_2 = \left( \tr X_{\Phi}^* X_{\Phi} \right)^{1/2} = \left( \sum_{i,j=1}^d  \sum_{k,l=1}^{d'} |C_{ij, kl}|^2 \right)^{1/2}
\ee

\medskip
Returning now to $\tr \Big[ B^* (\Phi \ot \Omega)(A) \Big]$, let $A \in {\cal M}_{d n}$ and
$B \in {\cal M}_{d' n'}$. Then we can write
\be
A = \sum_{i,j=1}^d | i \ket \bra j | \ot A_{ij}, \quad B = \sum_{k,l=1}^{d'} | k \ket \bra l | \ot B_{kl}
\ee
where in the first equation $\{ | i \ket \}$ is an orthonormal basis of $\mathbb{C}^d$ and $\{ A_{ij} \} \in {\cal M}_n$,
and in the second equation $\{ | k \ket \}$ is an orthonormal basis of $\mathbb{C}^{d'}$ and $\{ B_{kl} \} \in {\cal M}_{n'}$.
We have
\be
(\Phi \ot \Omega)(A) &=& \sum_{i,j=1}^d \Phi(| i \ket \bra j |) \ot \Omega(A_{ij}) \nonumber \\
&=& \sum_{i,j=1}^d \sum_{k,l=1}^{d'} C_{ij, kl} \, | k \ket \bra l | \ot \Omega(A_{ij})
\ee
and hence
\be\label{eq4}
\tr \Big[ B^* (\Phi \ot \Omega)(A) \Big] &=& \sum_{i,j=1}^d \sum_{k,l=1}^{d'} C_{ij, kl} \, \tr \Big(B_{kl}^* \Omega(A_{ij}) \Big)
\ee
Now we apply H\"older's inequality and the definition (\ref{purity-CP}) to get the bound
\be\label{ineq5}
| \tr \left(B_{kl}^* \Omega(A_{ij}) \right) | \le \| \Omega \|_{q \rightarrow p} \, \| B_{kl} \|_{p'} \, \| A_{ij} \|_q
\ee
where again $p'$ is the conjugate value defined by (\ref{def:p'}).
Combining (\ref{eq4}) and (\ref{ineq5}), and using the Cauchy-Schwarz inequality we get
\be\label{ineq6}
\Big| \tr \Big[ B^* (\Phi \ot \Omega)(A) \Big] \Big| & \le & \| \Omega \|_{q \rightarrow p} \, \sum_{i,j=1}^d \sum_{k,l=1}^{d'} | C_{ij, kl} | \, \| A_{ij} \|_q \,  \| B_{kl} \|_{p'}   \nonumber \\
& \le & \| \Omega \|_{q \rightarrow p} \, \left( \sum_{i,j=1}^d \sum_{k,l=1}^{d'} |C_{ij, kl}|^2 \right)^{1/2} \,
\left( \sum_{i,j=1}^d  \sum_{k,l=1}^{d'} {\| A_{ij} \|_q}^2 \, {\| B_{kl} \|_{p'}}^2 \right)^{1/2} \nonumber \\
& = & \| \Omega \|_{q \rightarrow p} \, \| X_{\Phi} \|_2 \,
\left( \sum_{i,j=1}^d {\| A_{ij} \|_q}^2 \right)^{1/2} \, \left( \sum_{k,l=1}^{d'} {\| B_{kl} \|_{p'}}^2 \right)^{1/2}
\ee

\medskip
We will use the following result of Bhatia and Kittaneh \cite{BhatKitt}: let $M \in {\cal M}_{ d n }$, written in the block form
\be
M = \sum_{i,j=1}^d | i \ket \bra j | \ot M_{ij}, \quad \{ M_{ij}\} \in {\cal M}_n
\ee
Then
\be\label{B-K1}
\sum_{i,j=1}^d {\| M_{ij} \|_q}^2 \le \begin{cases}{ \| M \|_q^2 & for $1 \le q \le 2$ \cr & \cr d^{2 - 4/q} \| M \|_q^2 & for $q \ge 2$ }\end{cases}
\ee
There are three separate cases to consider for the bound (\ref{ineq6}) depending on the values of $q,p$.

\medskip
\par\noindent \fbox{$q \le 2 \le p$}
In this case $q \le 2, p' \le 2$ so we can apply the first case in (\ref{B-K1}) to both sums in (\ref{ineq6}), and this gives the bound
\be
| \tr B^* (\Phi \ot \Omega)(A)  | & \le & \| \Omega \|_{q \rightarrow p} \, \| X_{\Phi} \|_2 \, \| A \|_q \,  \| B \|_{p'}
\ee
Applying (\ref{def:Holder-dual}) this leads to
\be
\| \Phi \ot \Omega \|_{q \rightarrow p} \le  \| \Omega \|_{q \rightarrow p} \, \| X_{\Phi} \|_2
\ee
and hence
\be\label{bd1}
\| \Phi \|_{q \rightarrow p}^{(pot)} = \sup_{n} \sup_{\Omega_n} \frac{\| \Phi \ot \Omega_n \|_{q \rightarrow p}}{\| \Omega_n \|_{q \rightarrow p}} \le \| X_{\Phi} \|_2
\ee

\medskip
\par\noindent \fbox{$2 \le q \le p$}
Since $q \ge 2, p' \le 2$ we apply the second case in (\ref{B-K1}) to $\sum_{i,j=1}^d {\| A_{ij} \|_q}^2$, and the first case to $\sum_{k,l=1}^{d'} {\| B_{kl} \|_{p'}}^2$.
This gives the bound
\be
| \tr B^* (\Phi \ot \Omega)(A)  | & \le & \| \Omega \|_{q \rightarrow p} \, \| X_{\Phi} \|_2 \, d^{1 - 2/q} \, \| A \|_q \,  \| B \|_{p'}
\ee
which directly implies the second case in (\ref{thm1:eq2}).

\medskip
\par\noindent \fbox{$q \le p \le 2$}
In this case $q \le 2, p' \ge 2$ so we use the second case in (\ref{B-K1}) for $\sum_{k,l=1}^{d'} {\| B_{kl} \|_{p'}}^2$, and the first case for $\sum_{i,j=1}^d {\| A_{ij} \|_q}^2$,
giving the bound
\be
| \tr B^* (\Phi \ot \Omega)(A)  | & \le & \| \Omega \|_{q \rightarrow p} \, \| X_{\Phi} \|_2  \, \| A \|_q \,  (d')^{1 - 2/p'} \, \| B \|_{p'}
\ee
Applying the definition (\ref{def:p'}) we arrive at the third case in (\ref{thm1:eq2}), and this completes the proof of Theorem \ref{thm1}.

\subsection{Proof of Theorem \ref{thm2}}
Again let $\Omega: {\cal M}_n \rightarrow {\cal M}_{n'}$ be a completely positive map, then
\be\label{thm2:pf:1}
\| \Phi \ot \Omega \|_{q \rightarrow p} = \sup_{A \in {\cal M}_{d n}} \frac{\| (\Phi \ot \Omega)(A) \|_p}{\| A \|_q}
\ee
Let $A \in {\cal M}_{d n}$, then we have
\be\label{ineq7}
\| (\Phi \ot \Omega)(A) \|_p & = & \| (\Phi \ot {\it id}_n) ({\it id}_d \ot \Omega)(A) \|_p \nonumber \\
& \le & \| \Phi \ot {\it id}_n \|_{p \rightarrow p} \, \| ({\it id}_d \ot \Omega)(A) \|_p \nonumber \\
& \le & \| \Phi \ot {\it id}_n \|_{p \rightarrow p} \, \| {\it id}_d \ot \Omega \|_{q \rightarrow p} \, \| A \|_q
\ee
Since $\Omega$ is completely positive and $q \le 2 \le p$ we may apply the result of Watrous \cite{Wat2005}
to deduce that
\be
\| {\it id}_d \ot \Omega \|_{q \rightarrow p} = \|  \Omega \|_{q \rightarrow p}
\ee
Furthermore the multiplicativity results from \cite{DJKR} imply that
\be
\| \Phi \ot {\it id}_n \|_{p \rightarrow p} = \| \Phi  \|_{p \rightarrow p} \, \| {\it id}_n \|_{p \rightarrow p} = \| \Phi  \|_{p \rightarrow p}
\ee
and hence from (\ref{ineq7}) and (\ref{thm2:pf:1}) we get
\be
\| \Phi \ot \Omega \|_{q \rightarrow p} \le \| \Phi  \|_{p \rightarrow p} \, \|  \Omega \|_{q \rightarrow p}
\ee
This implies the bound $\| \Phi \|_{q \rightarrow p}^{(pot)} \le  \| \Phi \|_{p \rightarrow p}$, and combining this with the first case
from Theorem \ref{thm1} we deduce the stated bound (\ref{thm2:eq1}).

\section{Proof of Theorem \ref{thm3}}\label{sect5}
Let $\Phi : {\cal M}_d \rightarrow {\cal M}_{d'}$ be a CQ map, so that its action on ${\cal M}_d$ has the form
\be\label{def:Phi-CQ}
\Phi(M) = \sum_{k=1}^d \bra k | M | k \ket \, R_k, \quad M \in {\cal M}_d
\ee
for some collection of positive semidefinite matrices $\{ R_k  \in {\cal M}_{d'}^{+}\}$. 
Let $\Omega : {\cal M}_n \rightarrow {\cal M}_{n'}$ be completely positive, then from (\ref{purity-CP}) we have
\be\label{thm3:pf:1}
\| \Phi \ot \Omega \|_{q \rightarrow p} = \sup_{A \in {\cal M}_{d n}^{+}} \frac{\| (\Phi \ot \Omega)(A) \|_p}{\| A \|_q}
\ee
So consider a matrix $A \in {\cal M}_{d n}^{+}$: we write
\be
A = \sum_{i,j=1}^d |i \ket \bra j | \ot A_{ij}, \quad \{A_{ij}\} \in {\cal M}_n
\ee
where the positivity of $A$ implies in particular that
$A_{ij}^* = A_{ji}$  and $A_{jj} \ge 0$, for all $i,j=1,\dots,d$. 
Then
\be\label{pf3:eq3}
(\Phi \ot \Omega)(A) = \sum_{k=1}^d R_k \ot \Omega(A_{kk}) \, \in \,{\cal M}_{d' n'}^{+}
\ee
We will prove Theorem \ref{thm3} by deriving a bound for $\| (\Phi \ot \Omega)(A) \|_p$.
For the proof we will assume that $A_{kk} \neq 0$ for all $k=1,\dots,d$, and hence that
$\| A_{kk} \|_q > 0$ (using a continuity argument it is easy to see
that it is sufficient to prove the bound under this assumption).

\medskip
Let $| \psi \ket$ be a unit vector in $\mathbb{C}^d$, and define
\be\label{def:B}
B = \sum_{k=1}^d R_k \ot | \psi \ket \bra \psi | \ot \Omega(A_{kk}) \in {\cal M}_{d' d n'}^{+}
\ee
Since $| \psi \ket \bra \psi |$ is a pure state, we have
\be\label{B=}
\| B \|_p = \| (\Phi \ot \Omega)(A) \|_p \quad \mbox{for all $p \ge 1$}
\ee
We follow a method introduced in \cite{King-EB1}, and note that
the expression (\ref{def:B}) can be factorized in the following way:
for any positive $\{x_1,\dots,x_d\}$,
\bee
B = (R \ot I_{n'}) \, W \, (R \ot I_{n'})^*
\eee
where
\be\label{def:R,x}
R = \sum_{j=1}^d \sqrt{x_j R_j}  \ot | \psi \ket \bra j |
\ee
and
\be
W = I_{d'} \ot  \sum_{k=1}^d | k \ket \bra k | \ot \Omega(x_k^{-1} A_{kk}) \in {\cal M}_{d' d n'}^{+}
\ee
We will select the value of $x_k$ to be
\be
x_k =  \| A_{kk} \|_q \quad (k=1,\dots,d)
\ee
Note that $R$ depends only on the map $\Phi$ and the input $A$, and 
$W$ depends only on the map $\Omega$ and the input $A$.
Using the  Lieb-Thirring inequality \cite{LiebTh} we get
\be\label{ineq10}
\tr \, \left(B^p \right) &=& 
\tr \Big( (R \ot I_{n'}) \, W \, (R \ot I_{n'})^* \Big)^p \nonumber \\
& \le &   \tr \Big((R \ot I_{n'})^* (R \ot I_{n'})\Big)^p \, W^p  \nonumber \\
& =& \tr \Big( (R^* R)^p \ot I_{n'} \Big) \, W^p \nonumber \\
& =& \tr \Big( (R^* R)^p \ot I_{n'} \Big) \Big(I_{d'} \ot  \sum_{k=1}^d | k \ket \bra k | \ot (\Omega(x_k^{-1} A_{kk}))^p \Big) \nonumber \\
& =& \sum_{k=1}^d \tr \Big[ (R^* R)^p (I_{d'} \ot  | k \ket \bra k |) \Big] \, \tr \Big(\Omega(x_k^{-1} A_{kk})\Big)^p
\ee
Using the definiton of $x_k$ in (\ref{def:R,x}) we have
\be
\| \Omega(x_k^{-1} A_{kk}) \|_p \le    \| \Omega \|_{q \rightarrow p} \, \| x_k^{-1} A_{kk} \|_q = \| \Omega \|_{q \rightarrow p}
\ee
and so (\ref{ineq10}) leads to
\be\label{ineq12}
\tr \, \left(B^p \right) &\le& 
\| \Omega \|_{q \rightarrow p}^p \, \sum_{k=1}^d \tr \Big[ (R^* R)^p (I_{d'} \ot  | k \ket \bra k |) \Big] \nonumber \\
&=& \| \Omega \|_{q \rightarrow p}^p \,  \tr  (R^* R)^p \nonumber \\
&=& \| \Omega \|_{q \rightarrow p}^p \,  \tr  (R R^*)^p \nonumber \\
&=& \| \Omega \|_{q \rightarrow p}^p \,  \tr  \Big(\sum_{j=1}^d x_j R_j  \ot | \psi \ket \bra \psi |\Big)^p \nonumber \\
&=& \| \Omega \|_{q \rightarrow p}^p \,  \tr  \Big(\sum_{j=1}^d x_j R_j \Big)^p
\ee
Define the matrix
\be\label{def:C}
C = \sum_{j=1}^d x_j \, | j \ket \bra j | = \sum_{j=1}^d \| A_{jj} \|_q  \, | j \ket \bra j |
\ee
Then
\be
\Phi(C) = \sum_{j=1}^d x_j R_j
\ee
so from (\ref{ineq12}) we get
\be\label{ineq13}
\| B \|_p &\le& \| \Omega \|_{q \rightarrow p} \,  \Big\| \sum_{j=1}^d x_j R_j \Big\|_p \nonumber \\
& =& \| \Omega \|_{q \rightarrow p} \, \| \Phi(C) \|_p \nonumber \\
& \le & \| \Omega \|_{q \rightarrow p} \, \| \Phi \|_{q \rightarrow p} \, \| C \|_q
\ee
Furthermore
\be
\tr \, C^q = \sum_{j=1}^d  \tr A_{jj}^q = \tr \Big( \sum_j | j \ket \bra j | \ot A_{jj} \Big)^q \le \tr A^q 
\ee
where $\sum_j | j \ket \bra j | \ot A_{jj}$ is the block diagonal projection of $A$ with respect to the computational basis.
Therefore we deduce
\be
\| B \|_p \le  \| \Omega \|_{q \rightarrow p} \, \| \Phi \|_{q \rightarrow p} \, \| A \|_q
\ee
and hence from (\ref{B=}) that
\be
\| \Phi \ot \Omega \|_{q \rightarrow p}  \le   \| \Omega \|_{q \rightarrow p} \, \| \Phi \|_{q \rightarrow p}
\ee
Since this holds for all $\Omega$ we immediately get  $\| \Phi \|_{q \rightarrow p}^{(pot)} \le \| \Phi \|_{q \rightarrow p}$.
The reverse inequality (\ref{pot-ineq0}) follows immediately from the definition, so we  deduce that equality holds.

\medskip
The above result holds for any CQ map $\Phi$, and for any values $q,p \ge 1$.
If we consider now a QC map $\Psi$, we note that the adjoint map $\Psi^*$ is CQ and also that
\be
\| \Psi \|_{q \rightarrow p} = \| \Psi^* \|_{p' \rightarrow q'}
\ee
where $p', q'$ are the conjugate values for $p,q$.
Thus applying the result for CQ maps we find for any completely positive map $\Omega$,
\be
\| \Psi \ot \Omega \|_{q \rightarrow p} &=& \| \Psi^* \ot \Omega^* \|_{p' \rightarrow q'} \\
&=& \| \Psi^*  \|_{p' \rightarrow q'} \, \| \Omega^* \|_{p' \rightarrow q'} \\
&=& \| \Psi  \|_{q \rightarrow p} \, \|  \Omega \|_{q \rightarrow p} 
\ee 
and this proves the result.

\section{Proof of Theorem \ref{thm4}}\label{sect6}
Recall that a Hadamard map acts by taking the Hadamard poduct with a positive semidefinite matrix.
More specifically, let $\{ | i \ket \}$ ($i=1,\dots,d$) be an orthonormal basis in $\mathbb{C}^d$, and let
$C = (c_{ij}) \in {\cal M}_d^{+}$ be a positive semidefinite matrix  so that
\be
C = \sum_{i,j=1}^d c_{ij} \, | i \ket \bra j |
\ee
Then we define the Hadamard map $H_C : {\cal M}_d \rightarrow {\cal M}_d$ by
\be
M = \sum_{i,j=1}^d m_{ij} \, | i \ket \bra j | \mapsto
H_C(M) = \sum_{i,j=1}^d c_{ij} \, m_{ij} \, | i \ket \bra j |
\ee
Let $\Omega : {\cal M}_n \rightarrow {\cal M}_{n'}$ be a completely positive
map, and $A \in {\cal M}_{d n}^{+}$ so that
\be
A =  \sum_{j,k=1}^d |j \ket \bra k | \ot A_{jk}
\ee
where $\{ A_{jk} \} \in {\cal M}_n$ ($j,k=1,\dots,d$). Then 
\be
(H_C \ot \Omega)(A) = \sum_{j,k=1}^d c_{jk} \, |j \ket \bra k | \ot \Omega(A_{jk}) \in {\cal M}_{d n'}^{+}
\ee
Our goal is to establish the following inequality: for all $q \le 2 \le p$, and all $A \in {\cal M}_{d n}^{+}$,
\be\label{ineq20}
\| (H_C \ot \Omega)(A) \|_p \le \| H_C \|_{q \rightarrow p} \, \| \Omega \|_{q \rightarrow p} \, \| A \|_q
\ee
The inequality (\ref{ineq20}) implies $\| H_C \ot \Omega \|_{q \rightarrow p} \le \| H_C \|_{q \rightarrow p} \, \| \Omega \|_{q \rightarrow p}$, and 
hence this will give us $\| H_C \|_{q \rightarrow p}^{(pot)} \le \| H_C \|_{q \rightarrow p}$.

\medskip
We will first establish a separate result, namely that the inequality (\ref{ineq20}) holds 
when $A$ is replaced by a matrix $B \in {\cal M}_{d n}$ 
which has the special form
\be\label{A-special1}
B = \sum_{j=1}^d | j \ket \bra m | \ot B_{j}, \quad \{ B_j \} \in {\cal M}_n, \quad j=1,\dots,d
\ee
where $| m \ket$ is one of the basis vectors in $\mathbb{C}^d$. Let us define
\be
Y = (H_C \ot \Omega)(B) = \sum_{j=1}^d c_{jm} \, |j \ket \bra m | \ot \Omega(B_{j})
\ee
then our immediate goal is to show that
\be\label{ineq20a}
\| Y \|_p \le \| H_C \|_{q \rightarrow p} \, \| \Omega \|_{q \rightarrow p} \, \| B \|_q
\ee
Note that
\be
Y^* Y = \sum_{j=1}^d | c_{jm} |^2 \, | m \ket \bra m | \ot \Omega(B_{j})^* \, \Omega(B_{j})
\ee
Since $p \ge 2$ we use convexity to deduce that
\be\label{ineq21}
\| Y \|_p^2 &=& \| Y^* Y \|_{p/2} \nonumber \\
&=& \Big\|  \sum_{j=1}^d | c_{jm} |^2 \,  \Omega(B_{j})^* \, \Omega(B_{j}) \Big\|_{p/2} \nonumber \\
& \le  & \sum_{j=1}^d | c_{jm} |^2 \,  \| \Omega(B_{j})^* \, \Omega(B_{j}) \|_{p/2} \nonumber\\
& = & \sum_{j=1}^d | c_{jm} |^2 \,  \| \Omega(B_{j}) \|_{p}^2 \nonumber \\
& \le & \| \Omega \|_{q \rightarrow p}^2 \, \sum_{j=1}^d | c_{jm} |^2 \,  \| B_j \|_{q}^2
\ee
Define the  matrix $g = \sum_{j=1}^d \| B_j \|_{q} \, | j \ket \bra m |  \in {\cal M}_d$ then
\be
H_C(g) =  \sum_{j=1}^d c_{jm} \, \| B_j \|_{q} \, | j \ket \bra m | 
\ee
Since $H_C(g)^* H_C(g)$ is a multiple of $| m \ket \bra m |$ and $p \ge 2$ we have
\be
\sum_{j=1}^d | c_{jm} |^2 \,  \| B_j \|_{q}^2 = \tr (H_C(g)^* H_C(g)) = \| H_C(g)^* H_C(g) \|_{p/2}
\ee
Therefore from (\ref{ineq21}) we get
\be\label{ineq21a}
\| Y \|_p^2 & \le & \| \Omega \|_{q \rightarrow p}^2 \, \| H_C(g)^* H_C(g) \|_{p/2} \nonumber \\
& = & \| \Omega \|_{q \rightarrow p}^2 \, \| H_C(g) \|_{p}^2 \nonumber \\
& \le & \| \Omega \|_{q \rightarrow p}^2 \, \| H_C \|_{q \rightarrow p}^2 \, \| g \|_q^2 \nonumber \\
& = & \| \Omega \|_{q \rightarrow p}^2 \, \| H_C \|_{q \rightarrow p}^2 \, \| g^* g \|_{q/2}
\ee
where we have introduced the Schatten anti-norm defined for $0 < t < 1$, and $M \ge 0$, by
\be\label{def:Schatten-anti}
\| M \|_t = \left( \tr M^t \right)^{1/t}
\ee
Again, since $g^* g$ is a multiple of $| m \ket \bra m |$, we have $\| g^* g \|_{q/2} = \tr ( g^* g )$ and thus (\ref{ineq21a}) gives
\be\label{ineq21b}
\| Y \|_p^2 & \le & \| \Omega \|_{q \rightarrow p}^2 \, \| H_C \|_{q \rightarrow p}^2 \, \tr ( g^* g ) \nonumber \\
& = & \| \Omega \|_{q \rightarrow p}^2 \, \| H_C \|_{q \rightarrow p}^2 \, \sum_{j=1}^d  \| B_j \|_{q}^2 \nonumber \\
& =& \| \Omega \|_{q \rightarrow p}^2 \, \| H_C \|_{q \rightarrow p}^2 \, \sum_{j=1}^d  \| B_j^* B_j \|_{q/2}
\ee
The anti-norm satisfies the following superadditivity property \cite{BourinHiai2011}: let $M_1,\dots,M_k$ be positive semidefinite, then for all $0 < t \le 1$,
\be\label{anti-norm}
\left\| \sum_{i=1}^k M_i \right\|_t \ge \sum_{i=1}^k \| M_i \|_t
\ee
Since $q < 2$ and $B_j^* B_j$ is positive semidefinite for $j=1,\dots,d$, we may apply (\ref{anti-norm}) and get
\be\label{ineq21c}
\| Y \|_p^2 & \le & \| \Omega \|_{q \rightarrow p}^2 \, \| H_C \|_{q \rightarrow p}^2 \,   \left\| \sum_{j=1}^d  B_j^* B_j \right\|_{q/2} \nonumber \\
& = & \| \Omega \|_{q \rightarrow p}^2 \, \| H_C \|_{q \rightarrow p}^2 \,   \left\| \sum_{j=1}^d  | m \ket \bra m | \ot B_j^* B_j \right\|_{q/2} \nonumber \\
& =& \| \Omega \|_{q \rightarrow p}^2 \, \| H_C \|_{q \rightarrow p}^2 \, \| B^* B \|_{q/2} \nonumber \\
& = & \| \Omega \|_{q \rightarrow p}^2 \, \| H_C \|_{q \rightarrow p}^2 \, \| B \|_{q}^2
\ee
and this implies the bound (\ref{ineq20a}). A similar argument shows that
(\ref{ineq20a}) also holds for matrices $B$ that can be written in the form
\be\label{A-special2}
B = \sum_{j=1}^d | m \ket \bra j | \ot B_{j}, \quad \{ B_j \} \in {\cal M}_n, \quad j=1,\dots,d
\ee
where again $ | m \ket$ is a fixed basis vector.

\medskip
In order to establish the bound (\ref{ineq20}) for $A \in {\cal M}_{d n}^{+}$ we  will use an inductive argument. The induction will be carried out in the size of the matrix $A$.
Specifically, for $1 \le a \le d$ we define the set of matrices
\be\label{def:S-a}
{\cal S}_a = \Big\{ A \in {\cal M}_{d n}^{+} \,\Big|\, A =  \sum_{j,k=1}^a |j \ket \bra k | \ot A_{jk}, \quad \{ A_{jk} \} \in {\cal M}_n \Big\}
\ee
Our proof of (\ref{ineq20a}) for the special class (\ref{A-special1}) with $m=1$ shows that the bound (\ref{ineq20}) holds for all matrices in ${\cal S}_1$.
For the induction step we will assume that (\ref{ineq20}) holds for all matrices in ${\cal S}_a$ for some $1 \le a \le d-1$, and then we will prove that it also holds for all
matrices in ${\cal S}_{a+1}$. The induction argument will then imply that (\ref{ineq20}) holds for all matrices in ${\cal S}_d = {\cal M}_{d n}^{+}$,
and this will complete the proof.

\medskip
We will use the following inequality \cite{King-gen-Hann}: suppose that $M$ is a positive semidefinite matrix,
which can be written in $2 \times 2$ block form as
\be
M = \bmx{ X & Y \cr Y^* & W }\emx \ge 0
\ee
Then for all $p \ge 2$,
\be\label{gen-Hann}
\| M \|_p = \left\| \bmx{ X & Y \cr Y^* & W }\emx \right\|_p \le \left\| \bmx{ \| X \|_p & \| Y \|_p \cr \| Y \|_p & \| W \|_p }\emx \right\|_p
\ee
and the inequality is reversed for $1 \le p \le 2$.
The inequality (\ref{gen-Hann}) was originally proved  for the case where $X,Y,W$ are all square matrices
of equal dimension. By padding with additional
rows and columns of zeros if necessary, the general case easily follows from this. It will be useful to reformulate 
(\ref{gen-Hann}) in the following way: suppose $M \in {\cal M}_m^{+}$ for some $m \ge 1$, and let $Q_1$ be an orthogonal projection 
on $\mathbb{C}^m$, so $Q_1^2 = Q_1^* = Q_1$. Let $Q_2 = I_m - Q_1$, and define $M_{ij} = Q_i M Q_j$ for
$i,j =1,2$. Then we have $M = M_{11} + M_{12} + M_{21} + M_{22}$ and (\ref{gen-Hann}) implies for $p \ge 2$
\be\label{gen-Hann2}
\| M_{11} + M_{12} + M_{21} + M_{22} \|_p \le \left\| \bmx{ \| M_{11} \|_p & \| M_{12} \|_p \cr \| M_{21} \|_p & \| M_{22} \|_p }\emx \right\|_p
\ee
with the reverse inequality for $1 \le p \le 2$.

\medskip
Now  fix $a$ satisfying $1 \le a \le d-1$ and assume that (\ref{ineq20}) holds for all matrices in ${\cal S}_a$. Define the orthogonal projections
\be
P_1 = \sum_{j=1}^a | j \ket \bra j |, \quad P_2 = I_d - P_1 = \sum_{j=a+1}^d | j \ket \bra j |
\ee
Let $A \in {\cal S}_{a+1}$, and write $A = E + F + F^* + G$ where
\be\label{A-block}
E = (P_1 \ot I_n) A (P_1 \ot I_n), \quad
F = (P_1 \ot I_n) A (P_2 \ot I_n), \quad
G= (P_2 \ot I_n) A (P_2 \ot I_n)
\ee
The map $H_C \ot \Omega$ respects the block decomposition, and so
\be
(H_C \ot \Omega)(A) = X + Y + Y^* + W
\ee
where
\be\label{B-block}
X &=& (H_C \ot \Omega)(E) = (P_1 \ot I_n) (H_C \ot \Omega)(A) (P_1 \ot I_n), \nonumber \\
Y &=& (H_C \ot \Omega)(F) = (P_1 \ot I_n) (H_C \ot \Omega)(A) (P_2 \ot I_n), \nonumber \\
W &=& (H_C \ot \Omega)(G) = (P_2 \ot I_n) (H_C \ot \Omega)(A) (P_2 \ot I_n)
\ee
Applying (\ref{gen-Hann2}) with $Q_i = P_i \ot I_n$ ($i=1,2$) and $m= n d$ we deduce that
\be\label{ineq22}
\| (H_C \ot \Omega)(A) \|_p \le \left\| \bmx{ \| X \|_p & \| Y \|_p \cr \| Y \|_p & \| W \|_p }\emx \right\|_p
\ee
Since $E \in {\cal S}_a$ and $X = (H_C \ot \Omega)(E)$ we can apply the inductive hypothesis to deduce
\be\label{X-p}
\| X \|_p  \le \| H_C \|_{q \rightarrow p} \, \| \Omega \|_{q \rightarrow p} \, \| E \|_q
\ee 
Furthermore both $F$ and $G$ are in the special class of matrices (\ref{A-special1}) (with $| m \ket = | a+1 \ket$) and so we also have
\be\label{Y,W-p}
\| Y \|_{p} &=& \| Y^* \|_p \le  \| H_C \|_{q \rightarrow p} \, \| \Omega \|_{q \rightarrow p} \, \| F \|_q \nonumber \\
\| W \|_p & \le &  \| H_C \|_{q \rightarrow p} \, \| \Omega \|_{q \rightarrow p} \, \| G \|_q
\ee
Furthermore the right side of (\ref{ineq22}) is monotone increasing in its diagonal entries, 
and thus
\be\label{ineq24}
\| (H_C \ot \Omega)(A) \|_p \le
 \| H_C \|_{q \rightarrow p} \, \| \Omega \|_{q \rightarrow p} \, 
\left\| \bmx{ \| E \|_q & z \cr z & \| G \|_q }\emx \right\|_p
\ee
where
\be
z = \frac{\| Y \|_p}{\| H_C \|_{q \rightarrow p} \, \| \Omega \|_{q \rightarrow p}}
\ee
Note that the right side of (\ref{ineq24}) 
is monotone increasing in $z$ within the interval $0 \le z \le \sqrt{\| E \|_q \, \| G \|_q}$.
Furthermore applying (\ref{Y,W-p}) we deduce that
\be
z \le \| F \|_q \le \sqrt{\| E \|_q \, \| G \|_q}
\ee
where the second inequality follows from non-negativity of the $2 \times 2$ matrix 
\be
\bmx{ \| E \|_q & \| F \|_q \cr \| F \|_q & \| G \|_q }\emx
\ee
which is itself a consequence of non-negativity of $A$. It follows that the right side of (\ref{ineq24}) increases when $z$ is replaced by
$\| F \|_q$, and thus we find
\be\label{ineq31}
\| (H_C \ot \Omega)(A) \|_p \le \| H_C \|_{q \rightarrow p} \, \| \Omega \|_{q \rightarrow p} \, 
\left\| \bmx{ \| E \|_q & \| F \|_q \cr \| F \|_q & \| G \|_q }\emx \right\|_p
\ee
Since $p > q$ the right side (\ref{ineq31}) can only increase when the $p$-norm is replaced by the $q$-norm, thus
\be
\| (H_C \ot \Omega)(A) \|_p & \le & \| H_C \|_{q \rightarrow p} \, \| \Omega \|_{q \rightarrow p} \, 
\left\| \bmx{ \| E \|_q & \| F \|_q \cr \| F \|_q & \| G \|_q }\emx \right\|_q \nonumber \\
& \le & \| H_C \|_{q \rightarrow p} \, \| \Omega \|_{q \rightarrow p} \, 
\left\|  E  +  F  +  F^* +  G  \right\|_q \nonumber \\
& \le & \| H_C \|_{q \rightarrow p} \, \| \Omega \|_{q \rightarrow p} \, \| A \|_q
\ee
where the second inequality follows from (\ref{gen-Hann2}) (reverse inequality since $q < 2$). This completes the proof of the induction step, and thus the argument is done.

\section{Discussion}\label{sect7}
We have introduced the notion of potential output purity $\| \Phi \|_{q \rightarrow p}^{(pot)}$ for a completely positive map 
$\Phi$, and explored some of its basic properties.
Theorems \ref{thm1} and \ref{thm2} provide upper bounds for the potential purity, and Theorems  \ref{thm3} and \ref{thm4} presents several classes of maps where this quantity
is shown to be equal to the (standard) output purity. It is known that there are channels which exhibit a gap between output purity and
potential output purity at $q=1$; this follows from the
existence theorems for violation of additivity of minimal output Renyi entropy \cite{HaydenWinter}, \cite{Hastings}, \cite{ASW}.
A continuity argument then implies that there are maps $\Phi$ for which
$\| \Phi \|_{q \rightarrow p}^{(pot)} > \| \Phi \|_{q \rightarrow p}$ at least for $q$ close to 1. 
The arguments for violation of additivity of minimal output Renyi entropy rely heavily on the 
fact that the minimum is achieved with a pure input state. Since this is generally not the case for $q>1$,
it is not obvious how those arguments can be extended beyond $q=1$.

\medskip
The potential output purity provides an upper bound for the regularized purity, as shown in (\ref{ineq-seq}), and
this may have applications in a variety of settings. As one example, the mixing time for a quantum channel semigroup can 
be estimated using the logarithmic Sobolev constant, and for product semigroups this can be estimated
using the potential purity. To illustrate this application consider the depolarizing channel 
$\Delta_{\lambda}$ on ${\cal M}_d$:
\be
\Delta_{\lambda}(A) = \lambda A + \frac{1-\lambda}{d} \, \tr (A) \, I_d
\ee
When $\lambda = e^{-t}$ we can write $\Delta_{\lambda} = e^{t {\cal L}}$ as a one-parameter semigroup, for which
the value of the log-Sobolev constant $\alpha_2$ is known \cite{KPT}:
\be
\alpha_2({\cal L}) = \frac{2(1 - 2/d)}{\log(d-1)}
\ee
However the log-Sobolev constant $\alpha_2({\cal L}^{(n)})$ is not known for the product semigroup 
$\Delta_{\lambda}^{\ot n} = e^{t {\cal L}^{(n)}}$, so it is desirable to find a good lower bound for $\alpha_2({\cal L}^{(n)})$.
As shown in \cite{KPT} and \cite{HFW}, this can be done by combining a uniform upper bound for $\| \Delta_{\lambda}^{\ot n} \|_{2 \rightarrow 4}$
with interpolation arguments. Here we note that the Choi matrix of
$\Delta_{\lambda}$ is entrywise positive for $\lambda > 0$, and hence we can apply the results from \cite{KNR}
to deduce that
\be
\| \Delta_{\lambda} \|_{2 \rightarrow 4}^{(pot)} = \| \Delta_{\lambda} \|_{2 \rightarrow 4}
\ee
Using (\ref{ineq-seq}) this implies that $\| \Delta_{\lambda}^{\ot n} \|_{2 \rightarrow 4} = \| \Delta_{\lambda} \|_{2 \rightarrow 4}^n$
for all $n \ge 1$. Combining known results for the single channel $\Delta_{\lambda}$ and  the methods from \cite{KPT} and \cite{HFW} leads to the bound
\be
\alpha_2({\cal L}^{(n)}) \ge \frac{1 - 2/d}{\log(3) \, \log(d-1) + 2 (1 - 2/d)}
\ee
which is somewhat tighter than the estimates presented in \cite{KPT} and \cite{HFW}.

\medskip
There are a number of interesting avenues to pursue. One concerns the question whether 
$\| {\cal E} \|_{q \rightarrow p}^{(pot)} = \| {\cal E} \|_{q \rightarrow p}$ for all entanglement breaking channels ${\cal E}$,
and for all $q,p \ge 1$. In Theorem \ref{thm3} we proved this for the case where ${\cal E}$ is either CQ or QC, but 
it would be striking if the result were not true for a general EB map. Another interesting question is whether  the 
maximum value $\| \Phi \|_{q \rightarrow p}^{(pot)}$ in (\ref{def:pot-pur}) is achieved for finite $n$, where
$n$ is the dimension of the ancilla space on which $\Omega_n$ operates. Such a result is known for
$\| \Phi \ot {\it id}_n \|_{1 \rightarrow p}$ \cite{Wat2005}, but again the proof relies on a pure input state and may not readily
extend to $q > 1$.

\section*{Acknowledgements}
Some of the results in this paper were presented at the workshop ``Probabilistic and Algebraic Methods in Quantum Information Theory'' held at
Texas A\&M University in July 2017, and the author thanks the organizers for the invitation to speak at the workshop.

{~~}

\end{document}